\documentclass[12pt]{article}
\usepackage{graphicx,amssymb,amsmath,empheq}
\usepackage{authblk}
\usepackage{amsthm,bm}
\usepackage{empheq}
\usepackage{subfig} % caution: not compatible with revtex4
\usepackage{soul,physics}
\usepackage{hyperref}
\hypersetup{colorlinks = true, linkcolor=black, citecolor=black, urlcolor=blue}
\usepackage{url}

\theoremstyle{plain}

\usepackage{tikz}
\usepackage{tikz-cd}
\usetikzlibrary{arrows}
\usetikzlibrary{intersections}
\usetikzlibrary{shapes.geometric}
\usetikzlibrary{decorations.pathmorphing, patterns,shapes}
\usetikzlibrary{decorations.markings}

\tikzset{
  mid arrow/.style={postaction={decorate,decoration={
        markings,
        mark=at position .575 with {\arrow{stealth}}
  }}},
  near arrow/.style={postaction={decorate,decoration={
        markings,
        mark=at position .275 with {\arrow{stealth}}
  }}},
  far arrow/.style={postaction={decorate,decoration={
        markings,
        mark=at position .800 with {\arrow{stealth}}
  }}},
  snake arrow/.style={fixed point arithmetic, decorate, decoration={snake,amplitude=2pt, segment length=11pt},postaction={decoration={markings,mark=at position 0.625 with {\arrow{stealth}}},decorate}},
}
\tikzset{
  baseline = -0.5ex,
  wavy/.style = {
    thick,
    decorate,
    decoration={snake,amplitude=2pt,segment length=5pt}},
  swavy/.style = {
    thick,
    decorate,
    decoration={snake,amplitude=1.2pt,segment length=3pt}},
  sdot/.style = {
    circle,
    draw=none,
    fill=black,
    minimum size=2.5pt,
    inner sep=0pt},
  bdot/.style = {
    circle,
    draw=none,
    fill=black,
    minimum size=4pt,
    inner sep=0pt},
  svertex/.style = {
    circle,
    draw=black,
    thick,
    fill=lightgray,
    minimum size=8pt,
    inner sep=1pt},
  bvertex/.style = {
    circle,
    draw=black,
    thick,
    fill=lightgray,
    minimum size=24pt},
  bvertexsmall/.style = {
    circle,
    draw=black,
    thick,
    fill=lightgray,
    minimum size=7pt},
  bvertexnormal/.style = {
    circle,
    draw=black,
    thick,
    fill=lightgray,
    minimum size=16pt},
  mmvertex/.style = {
    circle,
    draw=black,
    thick,
    fill=lightgray,
    minimum size=10pt,
    inner sep=1pt},
  mvertex/.style = {
    circle,
    draw=black,
    thick,
    fill=lightgray,
    minimum size=12pt,
    inner sep=1pt},
  dvertex/.style = {
    circle,
    draw=black,
    thick,
    fill=gray,
    minimum size=25pt}}

\usepackage[nosort]{cite}

\newcommand{\R}{{\mathrm{R}}}
\newcommand{\A}{{\mathrm{A}}}

\newcommand{\iu}{{i\mkern1mu}}
\newcommand*\diff{\mathop{}\!\mathrm{d}}

\newcommand{\VF}{\Upsilon}

\title{
Size Winding Mechanism beyond Maximal Chaos
}
\author[1]{Tian-Gang Zhou}
\author[1]{Yingfei Gu\thanks{guyingfei@gmail.com}}
\author[2,3]{Pengfei Zhang\thanks{pengfeizhang.physics@gmail.com}}
\affil[1]{\normalsize \it Institute for Advanced Study, Tsinghua University, Beijing, 100084, China}
\affil[2]{\normalsize \it Department of Physics, Fudan University, Shanghai, 200438, China}
\affil[3]{\normalsize \it Shanghai Qi Zhi Institute, AI Tower, Xuhui District, Shanghai 200232, China}
\date{\today}

\begin{document}
  \maketitle
  
  \begin{abstract}
   The concept of information scrambling elucidates the dispersion of local information in quantum many-body systems, offering insights into various physical phenomena such as wormhole teleportation. This phenomenon has spurred extensive theoretical and experimental investigations. Among these, the size-winding mechanism emerges as a valuable diagnostic tool for optimizing signal detection. In this work, we establish a computational framework for determining the winding size distribution in large-$N$ quantum systems with all-to-all interactions, utilizing the scramblon effective theory. We obtain the winding size distribution for the large-$q$ SYK model across the entire time domain. Notably, we unveil that the manifestation of size winding results from a universal phase factor in the scramblon propagator, highlighting the significance of the Lyapunov exponent. These findings contribute to a sharp and precise connection between operator dynamics and the phenomenon of wormhole teleportation.
  \end{abstract}
  \tableofcontents

\section{Introduction}

During chaotic unitary evolution, localized initial information within interacting many-body systems rapidly disseminates across the entire system—a phenomenon known as information scrambling \cite{Hayden:2007cs,Sekino:2008he}. In the Heisenberg picture, this scrambling process is elucidated through the growth of simple operators, typically quantified by the operator size distribution \cite{Roberts:2014isa,roberts2018operator,Qi:2018bje,GuOperatorSizeDistribution2022}, which is a probability distribution in a specific operator basis. Nevertheless, a comprehensive understanding of operator growth entails considering both the amplitude and the phase of the generic operator wavefunction. In particular, the phase is crucial for several fascinating and counterintuitive properties of quantum systems, including wormhole teleportation \cite{WoottersTeleportingUnknownQuantum1993a,Gao:2016bin,maldacenaDivingTraversableWormholes2017,Gao:2018yzk, QiEternalTraversableWormhole2018,JafferisTraversableWormholeTeleportation2021a, QiEternalTraversableWormhole2018, YaoManyBodyQuantumTeleportation2022,schuster2022operator, WalterQuantumGravityLab1_2023, WalterQuantumGravityLab2_2023a}, which has experimental investigation on the Google Sycamore processor with nine qubits \cite{SpiropuluTraversableWormholeDynamics2022}, indicating an exciting era of studying ``quantum gravity in lab''. Previous investigations have identified the size-winding mechanism as a promising candidate \cite{WalterQuantumGravityLab1_2023, WalterQuantumGravityLab2_2023a,SpiropuluTraversableWormholeDynamics2022}. In this proposal, to maximize the teleportation signal, the phase of the operator wavefunction should satisfy a specific pattern, as we will now elaborate.

 To be concrete, we focus on systems consisting $N$ Majorana Fermions, denoted as $\chi_{j}$ with $j=1,2,\ldots,N$, and governed by the Hamiltonian $H$. Any composite operator $\hat{O}$ can be expressed in the Majorana basis as $\hat{O} = \sum_{n} \sum_{j_1<j_2<\ldots<j_{n}} c_{j_1 j_2\ldots j_n} i^{\lfloor n/2 \rfloor} \chi_{j_1} \chi_{j_2} \ldots \chi_{j_n}$ \cite{roberts2018operator,Qi:2018bje,GuOperatorSizeDistribution2022}, following the convention $\{\chi_j, \chi_k \} = 2\delta_{jk}$. Here, $\lfloor \cdot \rfloor$ is the floor function, and the factor $\iu^{\lfloor n/2 \rfloor}$ ensures the Hermiticity of the basis operator. The coefficients $c_{j_1 j_2\cdots j_n}$ represent the amplitudes in the orthonormal operator basis, behaving like wave functions. We define the length $n$ of the Majorana string as the \textit{size} for this basis element. 
\begin{figure}
    \centering
    \includegraphics[width=0.6\linewidth]{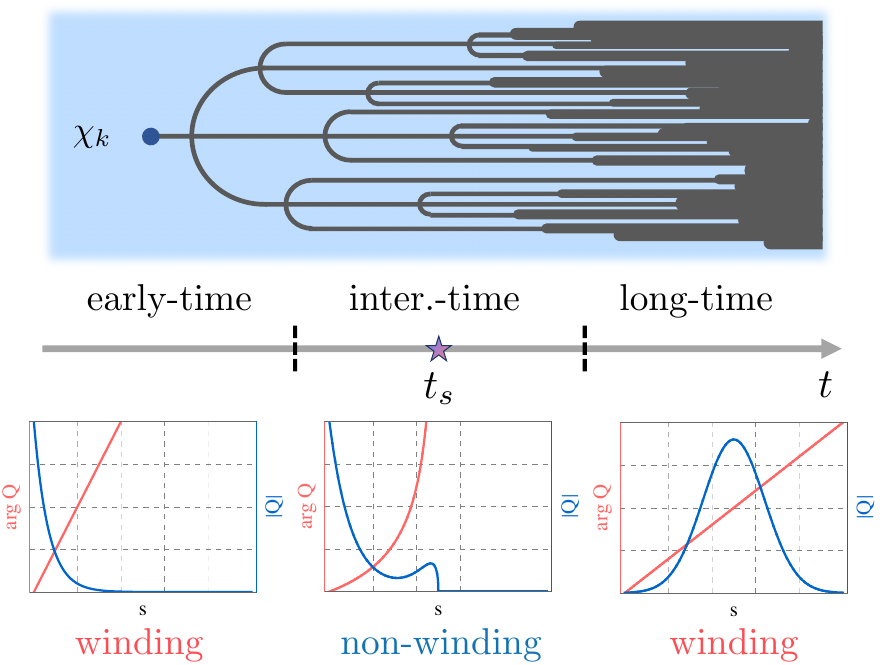}
    \caption{The schematics depict the operator growth at finite temperature. Our results unveil the manifestation of size winding in both the early-time regime and the long-time limit. } \label{fig1}
\end{figure}
In defining the size winding, we examine the operator $\hat{O}=\rho^{1/2}\chi_k(t)$, where $\rho$ is the thermal density matrix \cite{WalterQuantumGravityLab1_2023, WalterQuantumGravityLab2_2023a,schuster2022operator,SpiropuluTraversableWormholeDynamics2022}. The size winding, in its perfect form, refers to the situation where
\begin{equation}\label{eq:def_size_winding_size}
\text{Size winding:}\ \ \ \ \ \ c_{i_1,i_2,...,i_n}=e^{i(a+b n)}|c_{i_1,i_2,...,i_n}|.
\end{equation}
In other words, the phase of the operator wavefunction depends solely on the size $n$ through a linear function. We provide a concise review of the relationship between the teleportation signal and size winding in the Appendix \ref{app:A} for completeness. The size winding can also be probed by combining the standard size distribution $P(n,t)$ with the winding size distribution $Q(n,t)$, defined as:
    \begin{equation}\label{eq:size_winding_PQ}
        P(n,t) = \sum_{j_1<j_2<\cdots<j_{n}} \left|c_{j_1 j_2\cdots j_{n}}(t) \right|^2, \qquad
        Q(n,t) = \sum_{j_1<j_2<\cdots<j_{n}} \left(c_{j_1 j_2\cdots j_{n}}(t) \right)^2. 
    \end{equation}
The perfect size winding in Eq.~\eqref{eq:def_size_winding_size} is then equivalent to having $Q(n,t)/P(n,t)=e^{i(a+b n)}$. Known examples of systems with (near-)perfect size winding include the large-$q$ SYK model in the early-time regime and holographic systems with semi-classical gravity \cite{WalterQuantumGravityLab1_2023, WalterQuantumGravityLab2_2023a, schuster2022operator}. 

%We further introduce the concept of near-perfect size winding when $|Q(n, t)|$ is smaller than $P(n, t)$ by an $\mathcal{O}(1)$ factor. This reflects moderate phase fluctuations of $c_{j_1 j_2 \cdots j_{n}}$ for a large number of operators with the same size $n$. With this definition, systems exhibiting near-perfect size winding can also receive teleportation signals with $\mathcal{O}(1)$ amplitude, showing a sensitive dependence on the sign of the coupling strength.

In this work, we present a refined understanding of information scrambling by computing the size distribution $P(n,t)$ and the winding size distribution $Q(n,t)$ in large-$N$ quantum mechanics using the scramblon effective theory \cite{ZhangTwowayApproachOutoftimeorder2022, GuOperatorSizeDistribution2022, ZhangInformationScramblingEntanglement2023a,Liu:2023lyu}. We show that the scramblon propagator contributing to the generating function of $Q(n,t)$ acquires a pure imaginary time shift compared to $P(n,t)$, which is the origin of the size winding. This does not necessarily depend on maximal chaos, but rather relies on the universal chaotic behavior of the system \cite{larkin1969quasiclassical,almheiriApologiaFirewalls2013,shenkerBlackHolesButterfly2014,robertsDiagnosingChaosUsing2015,robertsLocalizedShocks2015,shenkerStringyEffectsScrambling2015,hosurChaosQuantumChannels2016a,maldacenaBoundChaos2016,stanfordManybodyChaosWeak2016a,maldacenaConformalSymmetryIts2016,maldacenaRemarksSachdevYeKitaevModel2016a,zhuMeasurementManybodyChaos2016,StanfordLocalCriticalityDiffusion2017,ReyMeasuringOutoftimeorderCorrelations2017,Sekino:2008he,Roberts:2014isa,Shenker:2014cwa,kitaev2014hidden}. Applying this approach to the large-$q$ SYK model \cite{YeGaplessSpinfluidGround1993,kitaevsimple,maldacena2016remarks,kitaev2018soft} yields the size winding distribution for the full range of time, thereby extending existing results \cite{WalterQuantumGravityLab1_2023, WalterQuantumGravityLab2_2023a, schuster2022operator}. The results (illustrated in FIG.~\ref{fig1}) reveal two different regimes for size winding. The first regime is the early-time regime $\varkappa^{-1}\ll t\ll t_{\text{sc}}$ with scrambling time $t_{\text{sc}}$ in the large-$N$ limit, where typical operators have $n\sim \mathcal{O}(1)$. The second regime corresponds to the long-time limit $t- t_{\text{sc}}\gg \varkappa^{-1}$, where  finite-$N$ corrections become crucial, indicating a completely different mechanism for size winding.

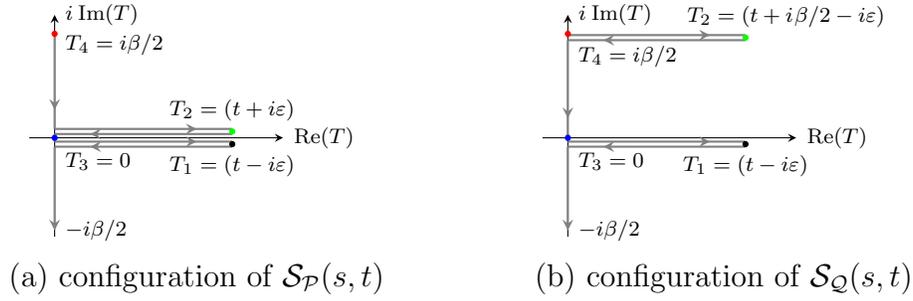
\begin{figure}[t]
    \centering
    \begin{tabular}{c@{\hspace{2.0cm}}c}
      \begin{tikzpicture}[scale=0.48,baseline={([yshift=-7pt]current bounding box.center)}]
        \draw [->,>=stealth] (-20pt,-57pt) -- (180pt,-57pt) node[right]{\scriptsize $\Re(T)$};
        \draw [->,>=stealth] (0pt, -135pt) -- (0pt,40pt) node[right]{\scriptsize $\iu \Im(T)$};
        \draw[thick,gray,far arrow] (0pt,25pt)--(0pt,-50pt);
        
        \draw[thick,gray,far arrow] (0pt,-50pt)--(140pt,-50pt);
        \draw[thick,gray,far arrow] (140pt,-54pt)--(0pt,-54pt);
        \draw[thick,gray] (0pt,-54pt)--(0pt,-60pt);
        \filldraw (140pt,-52pt) circle (0pt) node[above ]{\scriptsize$T_2=(t+i\varepsilon)$};
        \draw[thick,gray,far arrow] (0pt,-60pt)--(140pt,-60pt);
        \draw[thick,gray,far arrow] (140pt,-64pt)--(0pt,-64pt);
        \draw[thick,gray,->,>=stealth] (0pt,-64pt)--(0pt,-130pt);
        \filldraw (140pt,-62pt) circle (2pt) node[below ]{\scriptsize$T_1=(t-i\varepsilon)$};

        \filldraw (0pt,-60pt) circle (0pt) node[below right]{\scriptsize $T_3=0$};
        \filldraw (0pt,-130pt) circle (0pt) node[ right]{\scriptsize $-i\beta/2$};
        
        \filldraw (0pt,0pt) circle (0pt) node[above right]{\scriptsize $T_4=i\beta/2$};
        
        \filldraw[blue] (0pt,-57pt) circle (2pt);
        \filldraw[red] (0pt,25pt) circle (2pt);
        \filldraw[green] (140pt,-52pt) circle (2pt);
        
      \end{tikzpicture}
      &
      \begin{tikzpicture}[scale=0.48,baseline={([yshift=-7pt]current bounding box.center)}]
        \draw [->,>=stealth] (-20pt,-57pt) -- (180pt,-57pt) node[right]{\scriptsize $\Re(T)$};
        \draw [->,>=stealth] (0pt, -135pt) -- (0pt,40pt) node[right]{\scriptsize $\iu \Im(T)$};
        \draw[thick,gray,far arrow] (0pt,20pt)--(0pt,-57pt);
        
        \draw[thick,gray,far arrow] (0pt,24pt)--(140pt,24pt);
        \draw[thick,gray,far arrow] (140pt,20pt)--(0pt,20pt);
        \draw[thick,gray] (0pt,-54pt)--(0pt,-60pt);
        \filldraw (172pt,22pt) circle (0pt) node[above ]{\scriptsize$T_2=(t+\iu \beta /2 - i\varepsilon)$};
        \draw[thick,gray,far arrow] (0pt,-60pt)--(140pt,-60pt);
        \draw[thick,gray,far arrow] (140pt,-64pt)--(0pt,-64pt);
        \draw[thick,gray,->,>=stealth] (0pt,-64pt)--(0pt,-130pt);
        \filldraw (140pt,-62pt) circle (2pt) node[below ]{\scriptsize$T_1=(t-i\varepsilon)$};

        \filldraw (0pt,-60pt) circle (0pt) node[below right]{\scriptsize $T_3=0$};
        \filldraw (0pt,-130pt) circle (0pt) node[ right]{\scriptsize $-i\beta/2$};
        
        \filldraw (0pt,-10pt) circle (0pt) node[above right]{\scriptsize $T_4=i\beta/2$};
        
        \filldraw[blue] (0pt,-57pt) circle (2pt);
        \filldraw[red] (0pt,25pt) circle (2pt);
        \filldraw[green] (140pt,22pt) circle (2pt);
        
      \end{tikzpicture}
      \vspace{3pt}\\
      (a) configuration of $\mathcal{S}_\mathcal{P}(s,t)$ & (b) configuration of $\mathcal{S}_\mathcal{Q}(s,t)$
    \end{tabular}
    \caption{The Keldysh contour for the generating function of the operator size distribution is shown in the figure. The horizontal axes represent real-time, and the vertical axes represent imaginary time. The panels (a) and (b) correspond to the size distribution $\mathcal{S}_{\mathcal{P}}(s,t)$ and the winding size distribution $\mathcal{S}_{\mathcal{Q}}(s,t)$. The black, green, blue, and red dots label $T_1, T_2, T_3, T_4$, respectively.
    }
    \label{figcontourfinite}
  \end{figure}
\section{General Analysis}
To cover the full time range in which typical operators can get scrambled across the entire system, we normalize $s=n/N$ and consider the thermodynamic limit $N\rightarrow \infty$ \cite{GuOperatorSizeDistribution2022}. Consequently, $s \in [0,1]$ becomes a continuous variable and we further define continuous distributions $\mathcal{P}(s,t)=NP(sN,t)$ and $\mathcal{Q}(s,t)=NQ(sN,t)$. When computing distribution functions, it is more convenient to introduce generating functions through a Laplace transform. We have
\begin{equation}
 \mathcal{S}_{\mathcal{Q}}(\nu,t)=\int_0^1 \diff s~e^{-\nu s}\mathcal{Q}(s,t),
 \end{equation}
 and similarly for $\mathcal{S}_{\mathcal{P}}$. All moments of $s$ can be computed by taking derivatives with respect to $\nu$. After obtaining the generating functions, the size distribution and winding size distribution can be determined through an inverse Laplace transform.

\subsection{Generating functions}
Ref. \cite{Qi:2018bje} shows that the generating function of the size distribution can be naturally described by a correlation function in a double system. Subsequently, this connection has been extended to spin models \cite{Liu:2023lyu} and the winding size distribution \cite{WalterQuantumGravityLab1_2023, WalterQuantumGravityLab2_2023a, schuster2022operator}. We first introduce an auxiliary system with $N$ Majorana fermions $\psi_j$. We prepare the double system in the EPR state specified by $(\chi_j+i\psi_j)|\text{EPR}\rangle=0$, which is a vacuum state for complex fermion modes $c_j=(\chi_j+i\psi_j)/2$. Under this choice, applying a string operator of Majorana fermions $\chi_{j_1} \chi_{j_2} \ldots \chi_{j_n}$ to the EPR state creates $n$ complex fermions. Therefore, computing the size of operators is mapped to counting the particle number $\hat{n}=\sum_j c^\dagger_j c_j=\sum_j {(1+i\chi_j\psi_j)}/{2}$. This leads to 
\begin{equation}
\begin{aligned}
\mathcal{S}_\mathcal{P}(\nu,t)=\langle\text{EPR}|\rho^{1/2}\chi_k(t)e^{-\frac{\nu}{2N}\sum_j(1+i\chi_j\psi_j)}\chi_k(t)\rho^{1/2}|\text{EPR}\rangle,\\
\mathcal{S}_\mathcal{Q}(\nu,t)=\langle\text{EPR}|\rho^{1/2}\chi_k(t)e^{-\frac{\nu}{2N}\sum_j(1+i\chi_j\psi_j)}\rho^{1/2}\chi_k(t)|\text{EPR}\rangle,
\end{aligned}
\end{equation}
where we omit the arguments for conciseness. The generalization to spin models is straightforward using the size-total spin correspondence introduced in \cite{Liu:2023lyu}. 

Both generating functions possess a path-integral representation on the double Keldysh contour \cite{aleiner2016microscopic}. A closer examination reveals an important distinction in their imaginary time configurations, as illustrated in FIG~\ref{figcontourfinite}. Keeping to the first order in $1/N$ (reasons for going beyond the zeroth order are explained later), both $\mathcal{S}_\mathcal{P}(\nu,t)$ and $\mathcal{S}_\mathcal{Q}(\nu,t)$ can be expressed in a unified form
\begin{equation}\label{eq:generating_func}
\mathcal{S}_{\mathcal{P}/\mathcal{Q}}(\nu,t)=e^{-\frac{\nu}{2}+\frac{\nu^2}{8N}}\left< \text{T}_c \chi_k(T_1)\chi_k(T_2)\prod_j{\Big(1 + \frac{\nu~\chi_j(T_3)\chi_j(T_4)}{2N}\Big)}\right>.
\end{equation}
%\begin{equation}
%\mathcal{S}_{\mathcal{P}/\mathcal{Q}}(\nu,t)=\langle \text{T}_c \chi_k(T_1)\chi_k(T_2)e^{-\frac{\nu}{2}(1-\chi_j(T_3)\chi_j(T_4))}\rangle.
%\end{equation}
Here, $\text{T}_c$ represents the contour ordering operator. We introduce $T = t - \iu \tau$, where $t$ represents real-time and $\tau$ represents imaginary time. Both the traditional size distribution and the winding size distribution contain out-of-time-order (OTO) correlations \cite{larkin1969quasiclassical,Sekino:2008he,Roberts:2014isa,Shenker:2014cwa,kitaev2014hidden}. Therefore, the typical timescale for the evolution of the distribution functions is the scrambling time, and their calculation should involve recent developments in the scramblon effective theory \cite{ZhangTwowayApproachOutoftimeorder2022, GuOperatorSizeDistribution2022, ZhangInformationScramblingEntanglement2023a,Liu:2023lyu}. Moreover, in comparison to $\mathcal{S}_\mathcal{P}(\nu,t)$, there is a $\beta/2$ imaginary time shift for $T_2$ in the winding size generation function, which will give rise to the size winding phenomena.

\subsection{Scramblon calculation}

In the short-time limit specified by $\varkappa t \lesssim 1$, the evolution involves contributions from all microscopic details, making it non-universal. Therefore, our primary interest lies in the universal physics for $\varkappa t \gg 1$, where OTO-correlations dominate. In the scramblon effective theory, operators respect the Wick theorem unless they manifest OTO correlations, which are mediated by collective modes known as scramblons \cite{kitaev2018soft,Gu:2018jsv}. To warm up, a four-point OTO correlator (OTOC) can be computed by summing up diagrams where two pairs of operators interact by exchanging an arbitrary number of scramblons denoted by $m$:
  \begin{equation}\label{eq:OTOC}
  \begin{aligned}
    \langle\text{T}_c  {\chi}_j (T_1) {\chi}_j (T_2) &{\chi}_k (T_3) {\chi}_k (T_4) \rangle = \sum_{m=0}^\infty \frac{(-\lambda)^m}{m!} \VF^{\R,m} (T_{12}) \VF^{\A,m} (T_{34}),
    \end{aligned}
  \end{equation}
  where $\lambda = C^{-1} e^{\iu \varkappa \frac{\beta}{4}+ \varkappa \frac{T_1+T_2-T_3-T_4}{2}}$ is the propagator of the scrambling modes. The crucial phase factor $e^{\iu \varkappa {\beta}/{4}}$ ensures that the result is real for an equally spaced case with $\tau_1-\tau_3=\tau_2-\tau_4=\beta/4$. For the size generating function, configurations in FIG~\ref{figcontourfinite} corresponds to a real $\lambda=C^{-1}e^{\varkappa t}\equiv \lambda_0$ while for the winding size distribution, we find $\lambda=e^{i\varkappa \beta/4}\lambda_0$ due to the additional imaginary time shift. The prefactor $C$ is of the order of $N$ and, consequently, suppresses higher-order terms at an early time. We also introduce the scattering amplitude between fermions and $m$ scramblons in the future or past as $\VF^{\R,m}$ or $\VF^{\A,m}$.

  Applying a similar treatment, the generating functions $\mathcal{S}_{\mathcal{P}}$ and $\mathcal{S}_{\mathcal{Q}}$ in Eq.~\eqref{eq:generating_func} can be computed by summing over all configurations of scramblons. We expand the product of Majorana operators to the Majorana string basis, which leads to
        \begin{equation}\label{supp_eq:majo_expansion}
    \begin{split}
      \prod_{j=1}^N{\left(1+ \frac{\nu}{2N} \chi_j(T_3)\chi_j(T_4)\right)} &= \sum_{n=1}^{N} \sum_{|\bm{J}|=n} \left(  (-1)^{\frac{(n-1)n}{2}} 
         \chi_{\bm{J}}(T_3)\chi_{\bm{J}}(T_4) \left(\frac{\nu}{2N}\right)^{n}\right)   \\
    \end{split}.
  \end{equation}
    In the first summation, we consider $n$ as the length of Majorana strings, represented by $\chi_{\bm{J}} = \chi_{j_1} \chi_{j_2} \cdots \chi_{j_n}$, which ranges from 0 to $N$. The second summation encompasses all possible combinations of Majorana strings with length $|\bm{J}|=n$. Here, $\bm{J}$ denotes the indices of the Majorana strings, expressed as $\bm{J}\equiv j_1 j_2 \cdots j_n$. The indices $j_i$ are selected from the set ${1, \cdots, N}$, ensuring that $j_1\neq j_2 \cdots \neq j_n$. The inclusion of the extra phase $(-1)^{n(n-1)/2}$ arises from the anticommutation properties of Majorana fermions, particularly when combining pairs of Majorana fermions $\chi_j(T_3) \chi_j(T_4)$ into basis strings $\chi_{\bm{J}}(T_3) \chi_{\bm{J}}(T_4)$.

    Our approach to compute the generating function is analogous to the calculation of the OTOC as a summation of scramblon modes. In this case, we find
    \begin{equation}\label{eqn:newla}
        \begin{aligned}
            \mathcal{S}_{\mathcal{P}/\mathcal{Q}}(\nu,t) = &e^{-\frac{\nu}{2}+\frac{\nu^2}{8N}} \sum_{n=0}^N \sum_{|\bm{J}|=n} \sum_{m_1,\cdots,m_n} \VF^{\R,\sum_{l=1}^{n} m_l} (T_{12}) \\&\times\left( \frac{(-\lambda)^{m_1}}{m_1 !} \VF^{\A,m_1} (T_{34}) \cdots \frac{(-\lambda)^{m_n}}{m_n !} \VF^{\A,m_n} (T_{34}) \right) \left(\frac{\nu}{2N}\right)^n.
        \end{aligned}
    \end{equation}
    Here the summation over $\bm{J} = j_1, \cdots, j_n$ serves as a dummy variable, set to be evaluated when performing the ensemble average, similar to the process used in calculating the OTOC. Upon evaluation, the number of combinations for different indices $\bm{J}$, but of the same length $n$, is incorporated into $\sum_{|\bm{J}|=n}$. The resulting count of scramblon modes for a given $\bm{J}$ is denoted by $m_1, m_2, \cdots, m_n$ within the advanced vertex framework. Correspondingly, the scramblon mode count in the retarded vertex must align with the total number from all advanced vertices.

  Following the manipulations in Ref.~\cite{ZhangTwowayApproachOutoftimeorder2022}, the vertex functions $\VF^{\R,m}$ and $\VF^{\A,m}$ can be expressed as moments of $h^{\R/\A}$, i.e., $\VF^{\R/\A,m}(T)=\int_0^{+\infty} y^m h^{\R/\A}(y,T) \diff y$, which describe the distribution of the perturbation strength generated by Majorana fermion operators. To proceed, it is important to note that the retarded vertex in Eq. \eqref{eqn:newla} can be expressed through the function $h^R$ as
    \begin{equation}
        \VF^{\R,\sum_l m_l} (T_{12}) = \int_{0}^{\infty} \diff y y^{\sum_l m_l} h^R(y,T_{12}),
    \end{equation}
    and we can combine the introduced variable $y$ to scramblon propagator $\lambda$ and the advanced vertex to get another function $f^A$, which is defined as
    \begin{equation}
        \sum_l \frac{(-\lambda y)^{m_l}}{m_l !} \VF^{\A,m_l}(T_{34}) = \int_{0}^{\infty} \diff y_l e^{-\lambda y y_l}  h^A(y_l,T_{34}) = f^A(\lambda y, T_{34}).
    \end{equation}
    Physically, $f^A$ characterizes the fermion two-point function influenced by scramblon perturbations. Therefore, we arrive at
    \begin{equation}\label{supp_eq:S_gen_result}
        \begin{split}
            \mathcal{S}_{\mathcal{P}/\mathcal{Q}}(\nu,t) &= e^{-\frac{\nu}{2}+\frac{\nu^2}{8N}}  \int_{0}^{\infty} \diff y h^R(y,T_{12}) \sum_{n=0}^N \sum_{|\bm{J}|=n} \left( f^A(\lambda y, T_{34}) \right)^n \left(\frac{\nu}{2N}\right)^n \\
            &= e^{-\frac{\nu}{2}+\frac{\nu^2}{8N}}  \int_{0}^{\infty} \diff y h^R(y,T_{12})  \left(1 + f^A(\lambda y,T_{34}) \left(\frac{\nu}{2N}\right)\right)^N
        \end{split}
    \end{equation}
    Treating $\frac{\nu}{2N}$ as a small parameter, we expand the expression to the second order of $\left(\frac{\nu}{2N}\right)^2$ in the exponential term. We obtain closed-form results:
  \begin{equation}\label{eq:generating_func_res}
    \begin{split}
      \mathcal{S}_{\mathcal{P}/\mathcal{Q}}(\nu,t) &=  \int \diff y\, h^\R(y,T_{12})  
      e^{\frac{\nu^2}{8N}(1-f^A(\lambda y,-i\beta/2)^2) - \frac{\nu}{2} (1-f^A(\lambda y,-i\beta/2))},  \\
    \end{split}
  \end{equation}
  It describes the fermion two-point function under the influence of scramblon perturbation with strength $x$. By applying the inverse Laplace transform, we can approximate the size distribution or winding size distribution as
  \begin{equation}\label{eq:P_Q_distribution_finiteN}
    \begin{split}
      \mathcal{P}/\mathcal{Q}(s,t) &\approx \int_0^\infty \diff y~  \frac{h^\R(y,T_{12})}{\sqrt{2\pi \sigma^2}} e^{-\frac{1}{2 \sigma^2}\left(s - \frac{1-f^A(\lambda y,-i\beta/2)}{2}\right)^2},
    \end{split}
  \end{equation}
  with $\sigma^2 = {(1-f^A(\lambda y,-i\beta/2)^2)}/{4 N}$. Eq.~\eqref{eq:P_Q_distribution_finiteN} is the central result of this work, which is valid for generic chaotic large-$N$ quantum systems with all-to-all interactions, beyond models with maximal chaos. In the following sections, we analyze this result to unveil regimes which exhibits size winding.

  \section{Size winding at large-$N$}

  We can divide the the evolution into three time regimes: the early-time regime with $1 \ll \varkappa t \ll t_{\text{sc}}$, the intermediate time regime with $ t \approx t_{\text{sc}}$, and the long-time limit with $t- t_{\text{sc}}\gg \varkappa^{-1}$. In the first two regimes, the operator size distribution has an $\mathcal{O}(1)$ variance in continuous size $s$, allowing us to replace the Gaussian function in Eq.~\eqref{eq:P_Q_distribution_finiteN} with a delta function, as in the standard large-$N$ limit \cite{GuOperatorSizeDistribution2022}. In this case, the result can be expressed as
  \begin{equation}
  \begin{aligned}
  \mathcal{P}(s,t)&={2 }~{|\partial_y f^A(\lambda_0 y,-i\beta/2)|^{-1}}h^R(y,0),\\
  \mathcal{Q}(s,t)&={2 }~{|\partial_y f^A(\lambda_0 y,-i\beta/2)|^{-1}}h^R(e^{-i\varkappa \beta/4}y,-i\beta/2),
  \end{aligned}
  \end{equation}
  where $y$ is related to $s$ by solving $f^A(\lambda_0 y,-i\beta/2)=1-2s$. The validity of size winding mechanism can be examined by analyzing
  \begin{equation}
  \frac{\mathcal{Q}(s,t)}{\mathcal{P}(s,t)}=\frac{h^R(e^{-i\varkappa \beta/4}y,-i\beta/2)}{h^R(y,0)}.
  \end{equation}

  As demonstrated in Ref. \cite{gu2022two}, $h^R(y,-i\tau)$ is real and non-negative for real $y$ and arbitrary $\tau$. Therefore, the additional phase $e^{-i\varkappa \beta/4}$ directly contributes to the winding phase. The physical effect of a pure imaginary time shift has been previously discussed, particularly in the context of the chaos bound $\varkappa \leq \frac{2\pi}{\beta}$ \cite{maldacena2016bound}. The core of their argument lies in the observation that the out-of-time-ordered correlation function (OTOC), up to normalization, is (1) analytical and (2) less than or equal to 1 in the time region with a possible imaginary time shift $|\Delta \tau| \leq \beta / 4$. $\Delta \tau$ represents the imaginary time shift compared with the equally spaced configuration. A direct calculation reveals that the condition $\text{OTOC} \leq 1$ requires $\varkappa \leq \frac{2\pi}{\beta}$. Therefore, the physical significance of the imaginary time shift extends beyond the winding phase mechanism.
  
  \begin{figure}
    \centering
    \includegraphics[width=0.98\linewidth]{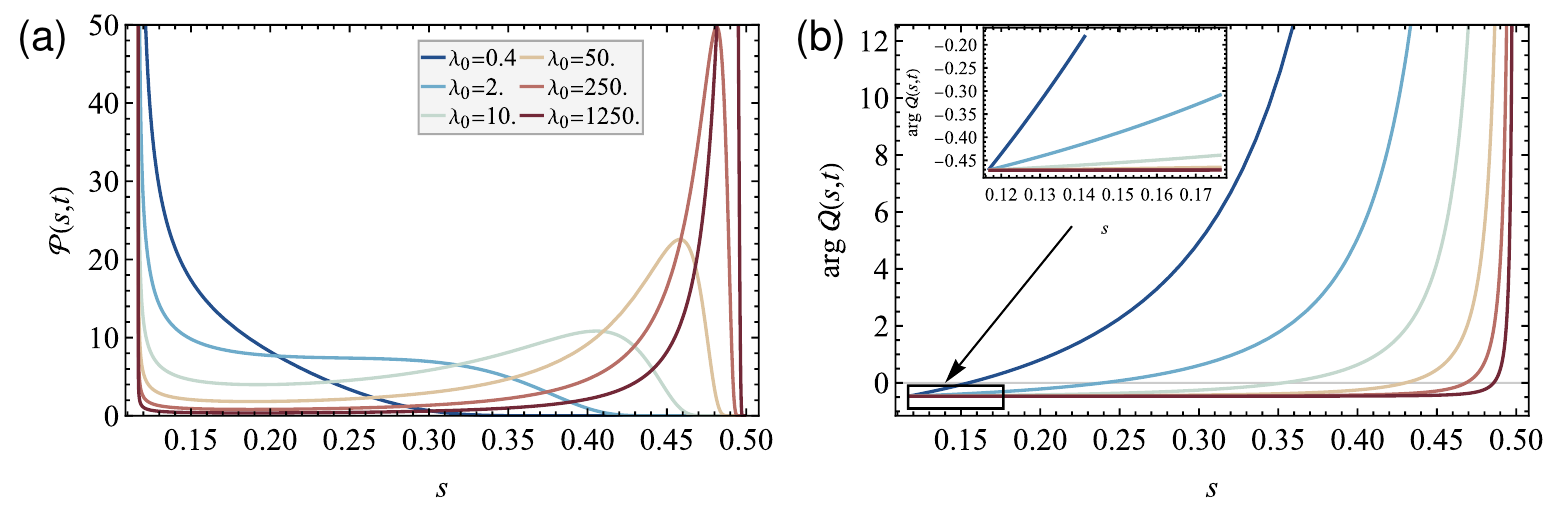}
    \caption{The analytical result of (a) $\mathcal{P}(s,t)$ and (b) $\arg \mathcal{Q}(s,t)$ extracted from Eq.~\eqref{eq:P_Q_result} with different $\lambda_0$ labeled in the legend. We choose the parameters as $\Delta=\frac{1}{4}$, $v=0.6$. In the inset of (b), we zoom in the $s\ll 1$ region and observe a linear dependence on size.}
    \label{fig:fig2windingsizeplot}
  \end{figure}

  To proceed, we examine an explicit example using the Sachdev-Ye-Kitaev model \cite{YeGaplessSpinfluidGround1993,kitaevsimple,maldacena2016remarks,kitaev2018soft}, which describes Majorana fermions with $q$-body random interactions. The Hamiltonian reads 
\begin{equation}\label{eq:Ham}
    H = \sum_{1 \le  j_1<j_2<\cdots<j_{q}\le N} J_{j_1\cdots j_q} \chi_{j_1} \chi_{j_2} \cdots \chi_{j_q},
  \end{equation}
  where $J_{j_1\cdots j_q}$ represents random Gaussian variables with a zero mean and a variance given by $\overline{J_{j_1\cdots j_q}^2}= \frac{(q-1)!\mathcal{J}^2}{2qN^{q-1}}$. We specifically focus on the large-$q$ limit by firstly taking $N\rightarrow \infty$ and then $q\rightarrow \infty$. In this scenario, both $h^R$ and $f^A$ have been computed in closed-form in Ref. \cite{gu2022two}:
\begin{equation}\label{supp_eq:HR_fA}
    \begin{split}
      h^R(y,T_{12}) &= \frac{y^{2 \Delta -1} \cos ^{2 \Delta }\left(\frac{\pi  v}{2}\right) \exp \left(- y \cos \left( \pi v \left(\frac{1}{2}- \frac{\iu T_{12}}{\beta}\right)  \right) \right)}{\Gamma (2 \Delta )}, \\
      f^A(\lambda y, T_{34}) &= \cos ^{2 \Delta }\left(\frac{\pi  v}{2}\right) \left(
      \cos \left( \pi v \left(\frac{1}{2}- \frac{\iu T_{34}}{\beta}\right)  \right)
      +\lambda  y\right)^{-2 \Delta }. \\
    \end{split}
  \end{equation}
We have introduced $\Delta=1/q$ and $\varkappa=2\pi v/\beta$ with $v = \frac{\beta \mathcal{J}}{\pi} \cos \frac{\pi v}{2}$. This leads to 
 \begin{equation}\label{eq:P_Q_result}
 \begin{aligned}
 %\mathcal{Q}(s,t) &= \frac{2 y^{2\Delta-1} \exp\left(- e^{-\frac{\iu\pi v}{2}} y\right) (1-2s)^{-\frac{2\Delta+1}{2\Delta}}}{\lambda_0 \Gamma(2\Delta+1)} e^{-\iu\Delta\pi v},\\
  \frac{\mathcal{Q}(s,t)}{\mathcal{P}(s,t)}&=\exp\bigg({i\sin\Big(\frac{\pi v}{2}}\Big)y-i\pi v\Delta\bigg).
  \end{aligned}
  \end{equation}
  Here, $y=\lambda_0^{-1}\left[\cos\Big(\frac{\pi v}{2}\Big)(1-2s)^{-\frac{1}{2\Delta}}-1\right]$, which indicates we have non-vanishing (winding) size distribution only for $s\in[\frac{1-\cos^{2\Delta}(\pi v/2)}{2},\frac{1}{2}]$. Interestingly, $\mathcal{P}(s,t) = |\mathcal{Q}(s,t)|$ is satisfied for arbitrary time. Nevertheless, the phase of $\mathcal{Q}(s,t)$ is linear in $y$ instead of $s$. Therefore, the large-$q$ SYK model exhibits perfect size winding only in the early-time regime $t\ll t_{\text{sc}}$ where typical operator has $s\ll 1$ and we can take the approximation $y\approx \lambda_0^{-1}\left[\cos\Big(\frac{\pi v}{2}\Big)(1+q s)-1\right]$. This is demonstrated by a numerical plot in FIG.~\ref{fig:fig2windingsizeplot}. We can further compute the slope of the winding phase as 
  \begin{equation}\label{eq:winding_phase_approx}
     \frac{\diff \arg \mathcal{Q}(s,t)}{\diff s} \approx  q\lambda_0^{-1}\sin \left({\pi  v}\right)/2 +\mathcal O(s^2).
  \end{equation}

  \section{Long-time limit \& finite $N$}
Now, we turn our attention to the long-time limit with $t- t_{\text{sc}}\gg \varkappa^{-1}$. The initial result Eq.~\eqref{eq:P_Q_result} naively suggests that the variance of the winding size distribution decays exponentially over time, eventually becoming smaller than the finite-$N$ broadening caused by a finite $\sigma$ in Eq.~\eqref{eq:P_Q_distribution_finiteN}. Furthermore, neglecting this finite-$N$ broadening leads to the divergence of $\text{arg}~Q$ near $s=1/2$, as illustrated in FIG.~\ref{fig:fig2windingsizeplot}. Consequently, to obtain the correct result, it is imperative to retain a finite $\sigma$ in Eq.~\eqref{eq:P_Q_distribution_finiteN}. This finite-$N$ correction is particularly crucial in the NISQ era, where quantum teleportation is performed on systems with a small number of qubits \cite{SpiropuluTraversableWormholeDynamics2022}. Additionally, it is essential for comparing theoretical predictions with numerics using exact diagonalization (ED), which is provided in the Appendix \ref{app:B}.

\begin{figure}
    \centering
    \includegraphics[width=1.0\linewidth]{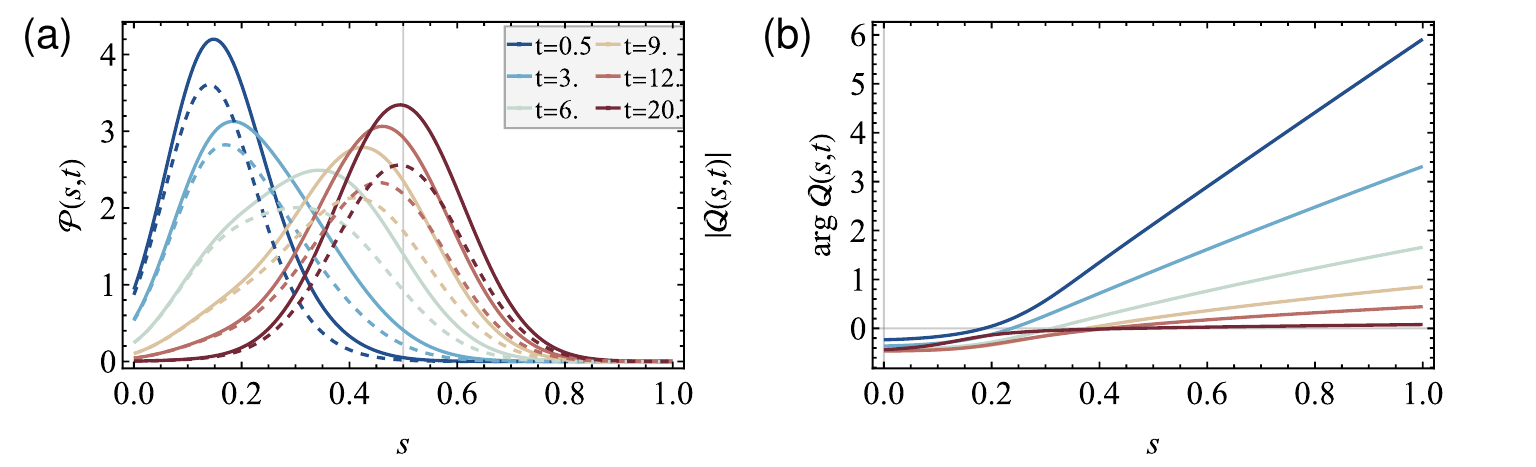}
    \caption{The size distribution and winding phase with finite-$N$ correction are numerically obtained from Eq.~\eqref{eq:P_Q_distribution_finiteN}. We use the parameters $\Delta=\frac{1}{4}$, $v=0.6$, $\beta=2\pi$, and $N=18$, motivated by the relatively small system size on NISQ devices. In (a), $\mathcal{P}(s,t)$ is represented by the solid line, and $|\mathcal{Q}(s,t)|$ is shown by the dashed line. In (b), $\arg \mathcal{Q}(s,t)$ represents the winding phase, and different colors are used to label different real-time $t$. }
    \label{fig:fig3windingsizefinitenplot}
  \end{figure}

  We numerically plot the size and winding size distributions with finite-$N$ corrections, as illustrated in FIG.~\ref{fig:fig3windingsizefinitenplot}. Firstly, we observe that $\mathcal{P}(s,t)$ and $|\mathcal{Q}(s,t)|$ are not exactly equal, although they qualitatively agree with each other, as seen in FIG.~\ref{fig:fig3windingsizefinitenplot}(a). $|\mathcal{Q}(n,t)|$ is smaller than $\mathcal{P}(n,t)$, indicating that the phase cancellation within fixed operator length sector appears as a leading finite-$N$ correction. Secondly, due to the finite-$N$ boardening, the peaks alway have finite widths and distribution functions are non-vanishing for the entire region of $s\in[0,1]$. Thirdly, the behavior of the winding phase with finite-$N$ correction undergoes significant changes, showing near-perfect size winding for arbitrary $s$ in the long-time limit, which was absent in the large-$N$ result. We can further expand Eq.~\eqref{eq:P_Q_distribution_finiteN} with $\lambda_0 \rightarrow \infty$ to obtain the asymptotic form of the winding phase. For the large-$q$ SYK model, this gives
  \begin{equation}\label{eq:argQ_finiteN}
    \begin{split}
      \arg \mathcal{Q}(s, t) &\approx (2 s-1) \frac{N}{\Gamma (2 \Delta )}   \left(\cos \left(\frac{\pi  v}{2}\right) e^{-\frac{2 \pi  t v}{\beta }}\right)^{2 \Delta } \\ &\left(\sin (\pi  \Delta  v) \left(\frac{2 \pi  t v}{\beta }-\psi ^{(0)}(2 \Delta )-2 \gamma \right)-\frac{1}{2} \pi  v \cos (\pi  \Delta  v)\right), \\
    \end{split}
  \end{equation}
  where $\psi^0(2\Delta)$ is the polyGamma function and $\gamma$ is the Euler constant. We find that this approximate formula matches well with Fig.~\ref{fig:fig3windingsizefinitenplot}(b) and confirms the linear behavior in the $s \sim \mathcal{O}(1)$ region. In Appendix \ref{app:B}, we have also employed the ED method to calculate the winding phase in the $N=18$ system, which also demonstrates linear behavior in the $s \sim \mathcal{O}(1)$ region. We thereby conclude that the $s \sim \mathcal{O}(1)$ linear winding phase receives important finite-size effects in small-size systems, indicating that teleportation operates via a completely different mechanism compared to the large-$N$ case.

\section{Discussion}
In this study, we examine the distribution of winding sizes as a detailed probe into information scrambling. Our results reveal that the winding phase is inherently linked to the pure imaginary time shift of an operator and its corresponding Keldysh contour. This connection results in a universal phase factor for scramblon propagators. The results show that the large-$q$ SYK model exhibits perfect size winding in the large-$N$ limit during the early-time regime, where typical operators possess a size $s \ll 1$. We further examine the effects of finite-$N$ corrections, which prove to be crucial in the long-time limit. This analysis uncovers a linear winding phase at $s \sim \mathcal{O}(1)$ in small-size systems, aligning with numerical simulations conducted through ED.

    Several additional remarks are pertinent. Firstly, although our focus is on chaotic quantum systems, recent studies have demonstrated that integrable systems, such as the commuting SYK model \cite{SpiropuluTraversableWormholeDynamics2022, gao2023commuting}, can also exhibit size winding in specific parameter regimes. Consequently, it becomes intriguing to inquire about the necessary conditions for the size winding mechanism.
    Secondly, while the holographic picture or maximal chaos are indispensable for the size winding mechanism, they can still play a crucial role in other facets of the wormhole teleportation protocol, such as the causal time-ordering of teleported signals and the Shapiro time delay. Thirdly, our results emphasize the pervasive influence of finite-$N$ effects in recent research, particularly when utilizing numerical tools or conducting experiments on NISQ quantum devices \cite{SpiropuluTraversableWormholeDynamics2022}. Advanced techniques are necessary to accurately observe and interpret theoretical results obtained in the large-$N$ limit.

\section*{Acknowledgment}
We thank Ping Gao, Xiao-Liang Qi, Jinzhao Wang and Ying Zhao for helpful discussions. The project is supported by NSFC under Grant No. 12374477 and the National Key R\&D Program of China 2023YFA1406702.

\appendix 
\section{Wormhole teleportation and size winding}\label{app:A}
In this section, we provide a brief review of the connection between the teleportation signal and size winding. To introduce the concept of winding size distribution, we consider the operator dynamics in a system comprising $N$ Majorana fermions. A composite operator $\hat{O}$ can be expanded using the Majorana basis, expressed as \[\hat{O} = \sum_{n} \sum_{j_1<j_2<\cdots<j_{n}} c_{j_1 j_2\cdots j_n} \iu^{\lfloor n/2 \rfloor} \chi_{j_1} \chi_{j_2} \cdots \chi_{j_n},\] adhering to the anti-commutation relation $\left\{\chi_j, \chi_k \right\} = 2\delta_{jk}$. Here, $\lfloor \cdot \rfloor$ denotes the floor function. The inclusion of $\iu^{\lfloor n/2 \rfloor}$ is crucial to maintain the Hermitian nature of the operator basis in the Majorana framework. The coefficients $c_{j_1 j_2 \cdots j_n}$, analogous to wave functions, represent the amplitudes on the corresponding basis operator. We define the 'size' of a Majorana operator string $\chi_{j_1} \chi_{j_2} \cdots \chi_{j_n}$ as its length $n$. We introduce two key distributions: the size distribution $P(n,t)$, representing the probability distribution, and the winding size distribution $Q(n,t)$:
    \begin{equation}\label{supp_eq:def_size_winding_size}
        P(n,t) = \sum_{j_1<j_2<\cdots<j_{n}} \left|c_{j_1 j_2\cdots j_{n}}(t) \right|^2, \ \ \ \ \ \ \ \ \ \
        Q(n,t) = \sum_{j_1<j_2<\cdots<j_{n}} \left(c_{j_1 j_2\cdots j_{n}}(t) \right)^2. 
    \end{equation}
  We focus on the time-dependent amplitudes $c_{j_1 j_2\cdots j_{n}}(t)$, which become significant when the composite operator $\hat{O}$ incorporates a Heisenberg-evolved operator $\hat{O}(t)$. Typically, $\hat{O}$ is non-Hermitian, resulting in complex amplitudes that can be expressed as $$c_{j_1 j_2\cdots j_{n}}(t) = |c_{j_1 j_2\cdots j_{n}}(t)| \exp(i \theta_{j_1 j_2\cdots j_{n}}(t)).$$ This complexity introduces a size-dependent phase in the winding size distribution, which we represent as $\operatorname{arg} Q(n,t) \equiv \theta(n,t)$. Previous studies have established that, ideally, $|Q(n,t)|$ is equivalent to $P(n,t)$, and $\theta(n,t)$ exhibits linear scaling with $n$. The linear scaling is indicative of optimal size winding, a crucial condition for achieving an optimal wormhole teleportation signal
 \cite{ YaoManyBodyQuantumTeleportation2022, WalterQuantumGravityLab1_2023, WalterQuantumGravityLab2_2023a,SpiropuluTraversableWormholeDynamics2022}.

    The process of wormhole teleportation naturally involves two-sided systems. For simplicity, we assume that the left and right systems are each constructed by $N$ Majorana Fermions $\chi_{j}^L, \chi_{j}^R, j=1,2,\cdots, N$. These left and right systems are governed by their corresponding Hamiltonians $H_L$ and $H_R$. Following the protocol in previous literature, the strength of the signal can be represented by a two-sided correlator\cite{maldacenaDivingTraversableWormholes2017,JafferisTraversableWormholeTeleportation2021a,YaoManyBodyQuantumTeleportation2022,WalterQuantumGravityLab1_2023,WalterQuantumGravityLab2_2023a,SpiropuluTraversableWormholeDynamics2022}
    \begin{equation}\label{supp_eq:two_side_corr}
      F(t) = \bra{\text{TFD}} e^{-igV} \chi_k^R(t) e^{igV} \chi_k^L(-t) \ket{\text{TFD}},
    \end{equation}
  where $V$ describes the coupling between the left and right system, and $g$ is the coupling strength. $\chi_k^L(-t)$ and $\chi_k^R(t)$ are the Heisenberg operators evolved by either $H_L$ or $H_R$. The $\ket{\text{TFD}}$ state is related to the EPR state by introducing the thermal density matrices $\rho_{\beta,L}=e^{-\beta H_L}$ and $\rho_{\beta,R}=e^{-\beta H_R}$, where $\ket{\text{TFD}} \equiv \rho^{1/2}_{\beta,L} \ket{\text{EPR}} = \rho^{1/2}_{\beta,R} \ket{\text{EPR}}$. This relation is ensured by the definition of the left and right Hamiltonians, $H_L = H_R^*$.
  
    We first note that in Eq. \eqref{supp_eq:two_side_corr}, the term $e^{-i g V}$ does not form an out-of-time-order correlator (OTOC) when combined with $\chi_k^R(t) \chi_k^L(-t).$ For sufficiently long time $t$, it can be approximately factored out. Therefore, the correlator can be reformulated as follows:
        \begin{equation}\label{supp_eq:corr-winding_size_distribution_step1}
    F(t) = e^{-\iu g \langle V \rangle} \bra{\text{EPR}} \rho_{\beta,R}^{1/2} \chi_k^R(t) e^{igV} \rho_{\beta,R}^{1/2} \chi_k^R(t) \ket{\text{EPR}}.
  \end{equation}
    The expectation of the coupling is defined under the TFD state. Secondly, by specifically choosing the coupling $V = \sum_j i \chi_j^L \chi_j^R,$ we can substantially simplify the expression of the correlator in Eq. \eqref{supp_eq:two_side_corr}. Up to a constant, the coupling operator $V$ matches the size operator, as discussed in the main text following Ref. \cite{qiQuantumEpidemiologyOperator2019}. Expanding $\rho_{\beta,R}^{1/2} \chi_k^R(t)$ in terms of Majorana fermion basis $\rho_{\beta,R}^{1/2} \chi_k^R(t) = \sum_{n} \sum_{j_1<j_2<\cdots<j_{n}} c_{j_1 j_2\cdots j_n} \iu^{\lfloor n/2 \rfloor} \chi_{j_1}^R \chi_{j_2}^R \cdots \chi_{j_n}^R$, the correlator can be related to the winding size distribution.
  \begin{equation}\label{supp_eq:corr-winding_size_distribution}
    F(t) = \sum_{n=0}^{N} e^{-\iu g \langle V \rangle} e^{\iu g (2 n - N )} Q(n,t),
  \end{equation} 
  where $Q(n,t)$ is the winding size distribution corresponding to the single side operator $\rho_{\beta}^{1/2}\chi_k(t)$.
  
  To attain the Maximal value of $F(t)$, two specific conditions are essential. The first condition is that the magnitude of $Q(n,t)$, denoted as $|Q(n,t)|$, should approximate $P(n,t)$, thereby placing constraints on the amplitude. The second condition entails setting $\arg Q(n,t)$ to $2i\alpha n$. This is intended to counterbalance the phase induced by the coupling, represented as $2ig n$. Implementing the relation $\alpha + g = 0$ and adhering to the normalization condition for the operator, which states that $\sum_n P(n,t) = 1$, we find that the two-sided correlator attains its peak value, $\max F(t) = e^{-ig (\langle V \rangle - N)}$. This demonstrates an optimal teleportation signal when the system exhibits perfect size winding \cite{YaoManyBodyQuantumTeleportation2022,WalterQuantumGravityLab1_2023,WalterQuantumGravityLab2_2023a,SpiropuluTraversableWormholeDynamics2022}.

    \section{Exact Diagonalization Numerics}\label{app:B}

\begin{figure}[t]
    \centering
        \includegraphics[width=1.\linewidth]{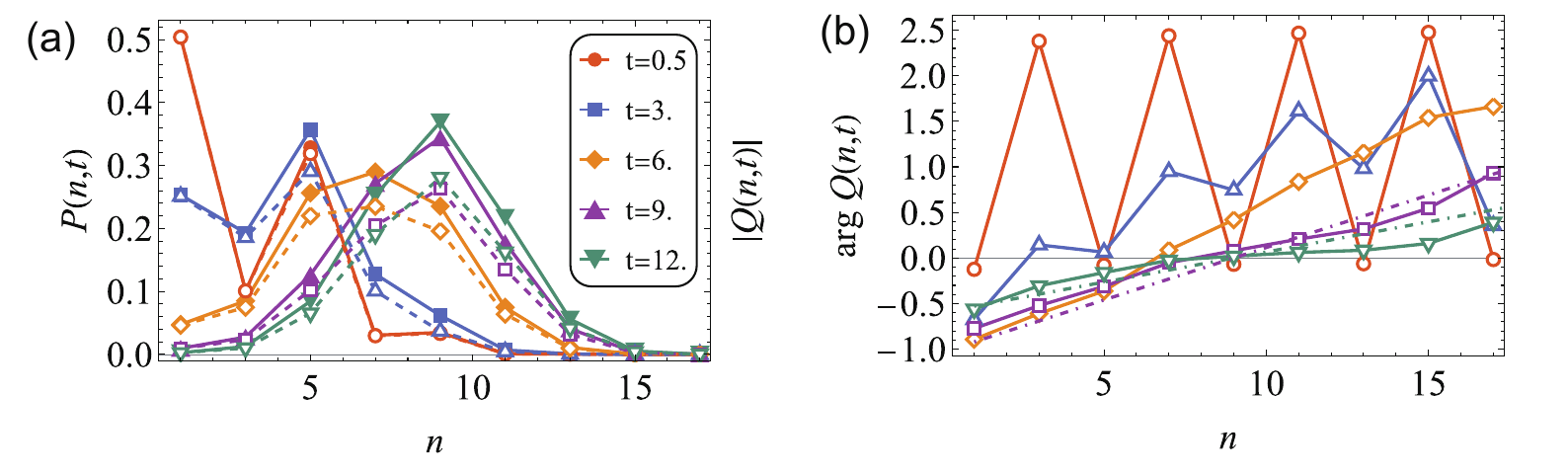}
    \caption{We present the results of exact diagonalization for (a) $P(n,t)$ and (b) $\arg Q(n,t)$, considering a Majorana fermion system with size $N=18$, under the normalization condition $\sum_n P(n,t)=1$. The chosen parameters are $\Delta=\frac{1}{4}$ and a temperature that corresponds to $v=0.6$. The result is averaged over 100 random realizations. In Fig. (a), $P(n,t)$ is depicted using solid lines and solid markers, whereas $|\mathcal{Q}(s,t)|$ is represented by dashed lines and open markers. In Fig. (b), $\arg \mathcal{Q}(s,t)$ is illustrated with solid lines and open markers, which denote the winding phase, and different colors signify various real-time instances $t$. The dot-dashed line corresponds to the leading-order approximation described in the main text (Eq.~\eqref{eq:argQ_finiteN}), applicable specifically for $t=9$ and $t=12$.}
    \label{fig:windingsize_ED}
  \end{figure}

    In this section, we provide the exact diagonalization results for size distribution and winding size distributions, as outlined in Eq.~\eqref{supp_eq:def_size_winding_size} for $q=4$ and $N=18$. Firstly, Fig.~\ref{fig:windingsize_ED} reveals that both $P(n,t)$ and $Q(n,t)$ have non-zero values only at odd operator string lengths $n$. This pattern is due to the conservation of the fermion parity. Secondly, the evolution of $P(n,t)$ and $Q(n,t)$ aligns qualitatively with the finite-$N$ results for a continuous variable of reduced size $s$, as shown in Fig.(3) of the main text. Initially, $P(n,t)$ and $|Q(n,t)|$ both peak around $n=1$, shifting to a peak at $n = 1/2 N$ in the later stages. Thirdly, $|Q(n,t)|$ consistently shows smaller values compared to $P(n,t)$. Early in the process, we observe notable fluctuations in both $P(n,t)$ and $|Q(n,t)|$ related to size $n$. These are not captured by the large-$q$ scramblon calculation in Fig.~\ref{fig:fig2windingsizeplot} of the main text. However, these fluctuations eventually decrease as time progresses.

    The winding phase depicted in Fig.~\ref{fig:windingsize_ED}(b) reveals a more complex structure in the context of real, finite-$N$ calculations. Initially, the early-time oscillations of the winding phase are governed by the anti-commutation properties of the fermion. These oscillations, however, are not apparent in the large-$q$ calculations that include finite-$N$ corrections. Subsequently, during the intermediate and late stages, the winding phase exhibits a linear trend across the entire operator size $n$ domain. Notably, at later times (e.g., $t=9, 12$), the winding phase can be qualitatively described by the approximation formula presented in the main text Eq.~\eqref{eq:argQ_finiteN}. This relationship is approximately proportional to $2n/N - 1$, with the slope diminishing over time. Consequently, our analysis suggests that the linear behavior of the winding phase in systems with small $N$ is predominantly influenced by finite-$N$ corrections, thereby overshadowing the large-$N$ behavior.
    
\bibliography{ref.bib}

\end{document}